# Observation of vector and tensor light shifts in $^{87}$Rb using near-resonant, stimulated Raman spectroscopy

Qing-Qing Hu[1,2,†], Christian Freier[2], Yuan Sun[1], Bastian Leykauf[2], Vladimir Schkolnik[2], Jun Yang[1], Markus Krutzik[2], Achim Peters[2]

[1]*Interdisciplinary Center for Quantum Information, National University of Defense Technology, Changsha, 410073, China*
[2]*Institut für Physik, Humboldt-Universität zu Berlin, 12489 Berlin, Germany*

**Abstract**

We present the derivation of the frequency dependent scalar, vector, and tensor dynamical polarizabilities for the two hyperfine levels of the $^{87}$Rb atom 5s ground state. Based on the characterization of the dynamical polarizabilities, we analyze and measure the differential vector and tensor light shift between the 5s ground state sub-levels with near-resonant, stimulated Raman transitions. These results clarify that the tensor polarizabilities for the ground states of alkali atoms are absent when the light field is far-detuned from the atomic resonance and the total electronic angular momentum J is a good quantum number. In the near resonant case, the light shifts are non-trivial and the determination of the frequency dependent vector and tensor dynamic polarizabilities will help to achieve higher fidelities for applications of neutral atoms in quantum information and precision measurements.

## I. Introduction

AC Stark shifts in alkali atoms, referring to the shift induced on a given energy level by the external oscillating electric fields, have been researched since 1960s [1]. For the frontiers of atomic physics, such as neutral atoms-based quantum information [2-5], the interactions between the atoms and the applied trapping beams inevitably cause light shifts on the atomic energy-levels. These light shifts are in general different for any two internal states of an atom and depend on the local intensity of the trapping beams [6,7], thus will affect the fidelity of the experiments via changing the detuning and Rabi frequency of the stimulated Raman transition. In particular, for qubits and quantum registers [2-4], the ground state Zeeman sub-levels other than the clock states are often employed, in which situation the vector and tensor light shift are non-trivial and will enlarge the effect of light shifts. For quantum precision measurements, such as atomic magnetometers [8-10], atomic clocks [11-13] and atomic interferometers [14-16], light shifts perturb the measurement by creating a fictitious magnetic field [17,18], or an undesired spatial variation and temporal fluctuation [13,19]. Especially for atom interferometry in a warm vapor [20], a significant light shift is created because the Raman laser Rabi frequency is up to 1.2 MHz. As a result, many efforts are currently focused on the measurement [17,18], elimination [18,21] and stabilization [22,23] of the light shifts. For example, these methods include hyper-Ramsey schemes [23-25], magic wavelength lattices [6,22], and careful adjustment the intensity ratio [19,26] or frequency [7,13] of the lasers.

The ac Stark shift is quantified using the dynamic polarizability, thus one has to know the exact values of the dynamic polarizabilities for the above purposes. For cesium atoms, the scalar, vector and tensor dynamical polarizabilities as well as the associated ac Stark shifts have been calculated and measured in Refs. [27,28]. In the case of rubidium, several measurements and theoretical calculations for the polarizabilities of the 5s ground state and few excited states have been carried out [12,17,18,21,29]. In particular, Ref. [12] provides a detailed calculation for the differential ac Stark shifts between the $^{87}$Rb atom 5s and 5p states, as well as an overview of the related experimental and theoretical work. In this paper, we evaluate the frequency dependent dynamical polarizabilities of the $^{87}$Rb atom 5s ground state sub-levels interacting with a pair of near-resonant Raman lasers, derive the differential vector and tensor light shifts between the two sub-levels, and verify the dynamical polarizabilities and differential light shifts experimentally. Section II presents the basic theory and calculated results. Section III presents the experimental setup used to measure the light shifts and the comparison between the inferred and measured results. Section IV gives a summary and an outlook.

## II. Theory
### A. AC Stark shift on a hyperfine state

Fig. 1 shows the D2 line hyperfine levels of $^{87}$Rb atoms and its Raman transitions in circularly polarized laser configurations in the presence of Zeeman and light shifts, where $F$ is the total angular momentum and $m_F$ is its projection on the magnetic quantization axis. The frequency difference of the two Raman lasers driving the Raman transition satisfies:

$$\hbar(\omega_2 - \omega_1) = \hbar\omega_{hfs} + \Delta E_{Zeeman} + \Delta E^{ac} \tag{1}$$

where $\omega_{hfs} \approx 2\pi \times 6.835 \text{GHz}$ is the ground state hyperfine splitting, $\hbar$ is the reduced Planck's constant, $\Delta E_{Zeeman} = \mu_B g_F m_F B$ is the first-order Zeeman shift, $\mu_B$ and $g_F$ are the Bohr magneton and Landé g factor respectively, $\Delta E^{ac}$ is the differential ac Stark shift between $|F=2, m_F\rangle$ and $|F=1, m_F\rangle$ ($m_F = 0, \pm 1$) sub-levels.

---

[†] Corresponding author. E-mail: qingqinghu@nudt.edu.cn; qingqinghu@physik.hu-berlin.de

For neutral atoms, whose ground state permanent electric dipole moment is zero, the shift on the $v$th energy level is quadratic dependent on E-fields as [13]:

$$\Delta E_v^{ac} = -\alpha_v(\omega)\left(\frac{\varepsilon}{2}\right)^2 \tag{2}$$

where the proportional coefficient $\alpha_v(\omega)$ is the called the dynamic polarizability, $\varepsilon$ and $\omega$ are the amplitude and frequency of the laser field $\boldsymbol{E}(t) = \varepsilon \hat{\zeta} e^{-i(\omega t + \hat{k}\cdot\hat{r})} + c.c.$. Here $c.c.$ represents the complex conjugate of the preceding term, $\hat{k}$ is the wave vector of the laser field, $\hat{\zeta}$ is the complex polarization vector of the laser and may be expressed as $\hat{\zeta} = e^{i\gamma}\left(\cos\varphi\hat{\zeta}_{maj} + i\sin\varphi\hat{\zeta}_{min}\right)$, in which $\gamma$ is a real number, the real unit vectors $\hat{\zeta}_{maj}$ and $\hat{\zeta}_{min}$ ($\hat{\zeta}_{maj} \times \hat{\zeta}_{min} = \hat{k}$) align with the semi-major axis and semi-minor axis of the ellipse which swept out by the electric field vector of the laser in one period, respectively [30]. $\varphi$ is defined by $\tan\varphi = |\hat{\zeta}_{min}|/|\hat{\zeta}_{maj}|$ and directly related to the degree of circular polarization A of the laser. For circularly polarized laser $\sigma^\pm$, $A = \pm 1$; for linearly polarized laser, $A = 0$.

The dynamic polarizability in Eq. (2) is defined as [12,13,31]:

$$\alpha_v(\omega) = \sum_{k\neq v}\frac{|\langle v|\boldsymbol{d}\cdot\hat{\zeta}|k\rangle|^2}{E_k - E_v - \hbar\omega} + \sum_{k\neq v}\frac{|\langle v|\boldsymbol{d}\cdot\hat{\zeta}|k\rangle|^2}{E_k - E_v + \hbar\omega} \tag{3}$$

where $E_k$ and $E_v$ represent the energies of atomic states $|k\rangle$ and $|v\rangle$, $\boldsymbol{d}$ is electric-dipole operator, and the sums are over a complete set of atomic eigenstates. For a particular hyperfine state $|F, m_F\rangle$, the dynamic polarizability is usually expressed in a form with decomposition into $m_F$ dependent factors and $m_F$ independent factors as [12,13,31]:

$$\alpha_{F,m_F}(\omega) = \alpha_F^S(\omega) + (\hat{k}\cdot\hat{B})A\frac{m_F}{2F}\alpha_F^V(\omega) + \left(3|\hat{\zeta}\cdot\hat{B}|^2 - 1\right)\frac{3m_F^2 - F(F+1)}{2F(2F-1)}\alpha_F^T(\omega) \tag{4}$$

where $\hat{B}$ is the unit vector along the quantization magnetic field. The $m_F$ independent factors $\alpha_F^S(\omega)$, $\alpha_F^V(\omega)$ and $\alpha_F^T(\omega)$, named as scalar, vector and tensor polarizabilities, respectively, are defined as [31]:

$$\alpha_F^S(\omega) = \sum_{F'}\frac{2\omega_{F'F}|\langle F\|\boldsymbol{d}\|F'\rangle|^2}{3\hbar(\omega_{F'F}^2 - \omega^2)} \tag{5}$$

$$\alpha_F^V(\omega) = \sum_{F'}(-1)^{F+F'+1}\sqrt{\frac{6F(2F+1)}{F+1}}\begin{Bmatrix}1 & 1 & 1\\ F & F & F'\end{Bmatrix}\frac{\omega_{F'F}|\langle F\|\boldsymbol{d}\|F'\rangle|^2}{\hbar(\omega_{F'F}^2 - \omega^2)} \tag{6}$$

$$\alpha_F^T(\omega) = \sum_{F'} (-1)^{F+F'} \sqrt{\frac{40F(2F+1)(2F-1)}{3(F+1)(2F+3)}} \begin{Bmatrix} 1 & 1 & 2 \\ F & F & F' \end{Bmatrix} \frac{\omega_{F'F} |\langle F \| d \| F' \rangle|^2}{\hbar(\omega_{F'F}^2 - \omega^2)} \quad (7)$$

where $\omega_{F'F}$ represents the $^{87}$Rb atom D2 line resonant frequencies of the hyperfine dipole transitions from ground state $F$ to excited state $F'$, namely the resonant frequencies of the $|F=1\rangle \to |F'=0,1,2\rangle$ and $|F=2\rangle \to |F'=1,2,3\rangle$ transitions without Zeeman shift and light shift, $\begin{Bmatrix} 1 & 1 & 1 \\ F & F & F' \end{Bmatrix}$ and $\begin{Bmatrix} 1 & 1 & 2 \\ F & F & F' \end{Bmatrix}$ are the Wigner 6-j symbols [31], $|\langle F \| d \| F' \rangle|$ is the reduced matrix element [32].

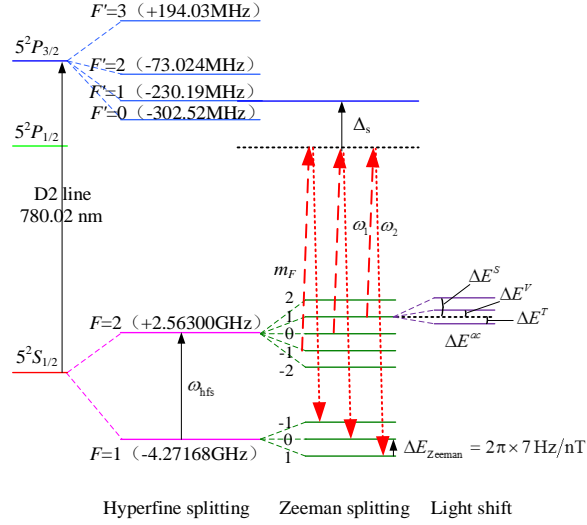

Fig. 1. $^{87}$Rb atom D2 line hyperfine levels and its Raman transition schemes in circularly polarized laser configurations in the presence of Zeeman and light shifts.

### B. Frequency dependent dynamical polarizabilities

By taking the resonant frequencies $\omega_{F'F}$ from Ref. [32], calculating the Wigner 6-j symbols and the reduced matrix elements in Eqs. (5)-(7), the frequency dependent scalar, vector and tensor polarizabilities $\alpha_F^l(\omega)$ ($l=S,V,T$) of the $^{87}$Rb atom $F=1$ and $F=2$ ground states are simplified to:

$$\alpha_1^S(\omega) = \left[ \frac{\omega_{01}}{9(\omega_{01}^2 - \omega^2)} + \frac{5\omega_{11}}{18(\omega_{11}^2 - \omega^2)} + \frac{5\omega_{21}}{18(\omega_{21}^2 - \omega^2)} \right] \frac{d_2^2}{\hbar}$$
$$\alpha_2^S(\omega) = \left[ \frac{\omega_{12}}{30(\omega_{12}^2 - \omega^2)} + \frac{\omega_{22}}{6(\omega_{22}^2 - \omega^2)} + \frac{7\omega_{32}}{15(\omega_{32}^2 - \omega^2)} \right] \frac{d_2^2}{\hbar} \quad (8)$$

$$\alpha_1^V(\omega) = \left[ -\frac{\omega_{01}}{6(\omega_{01}^2 - \omega^2)} - \frac{5\omega_{11}}{24(\omega_{11}^2 - \omega^2)} + \frac{5\omega_{21}}{24(\omega_{21}^2 - \omega^2)} \right] \frac{d_2^2}{\hbar}$$
$$\alpha_2^V(\omega) = \left[ -\frac{\omega_{12}}{20(\omega_{12}^2 - \omega^2)} - \frac{\omega_{22}}{12(\omega_{22}^2 - \omega^2)} + \frac{7\omega_{32}}{15(\omega_{32}^2 - \omega^2)} \right] \frac{d_2^2}{\hbar} \quad (9)$$

$$\alpha_1^T(\omega) = \left[ -\frac{\omega_{01}}{9(\omega_{01}^2 - \omega^2)} + \frac{5\omega_{11}}{36(\omega_{11}^2 - \omega^2)} - \frac{\omega_{21}}{36(\omega_{21}^2 - \omega^2)} \right] \frac{d_2^2}{\hbar}$$
$$\alpha_2^T(\omega) = \left[ -\frac{\omega_{12}}{30(\omega_{12}^2 - \omega^2)} + \frac{\omega_{22}}{6(\omega_{22}^2 - \omega^2)} - \frac{2\omega_{32}}{15(\omega_{32}^2 - \omega^2)} \right] \frac{d_2^2}{\hbar}$$
(10)

respectively, where $d_2 = |\langle J=1/2 \| er \| J'=3/2 \rangle|$ is the reduced D2 transition dipole matrix element of $^{87}$Rb.

Based on Eqs. (8)-(10), the frequency dependent scalar, vector and tensor polarizabilities $\alpha_F^S(\omega)$, $\alpha_F^V(\omega)$ and $\alpha_F^T(\omega)$ of the $^{87}$Rb atom ground state $F=1$ and $F=2$ sub-levels induced by a pair of near-resonant Raman laser fields $\mathbf{E}_1(\omega_1)$ and $\mathbf{E}_2(\omega_2)$ are shown in Figs. 2(a)-2(c), respectively, where $\omega_1 = \omega_{12} + \Delta_s$, $\omega_2 = \omega_1 + 6.834\,\text{GHz}$, $|\Delta_s| \leq 2\,\text{GHz}$. Fig. 2(d) shows the tensor polarizabilities between $-0.8\,\text{GHz}$ and $-0.6\,\text{GHz}$. Table 1 gives the specific polarizabilities when the single photon detuning is $\Delta_s = -700\,\text{MHz}$. Fig. 2 and Table 1 both show that the polarizabilities $\alpha_1^l(\omega_2)$ (red solid lines) and $\alpha_2^l(\omega_1)$ (blue dash lines) are much larger than the polarizabilities $\alpha_1^l(\omega_1)$ (red dash dot lines) and $\alpha_2^l(\omega_2)$ (blue dot lines) because in the latter cases the detunings are much larger ($\sim 6.8\,\text{GHz}$). Due to the difference between the dynamic polarizabilities of $F=1$ and $F=2$ sub-levels, a typical differential light shift will appear in Raman transition unless the intensities of the two Raman lasers are subtly chosen [33].

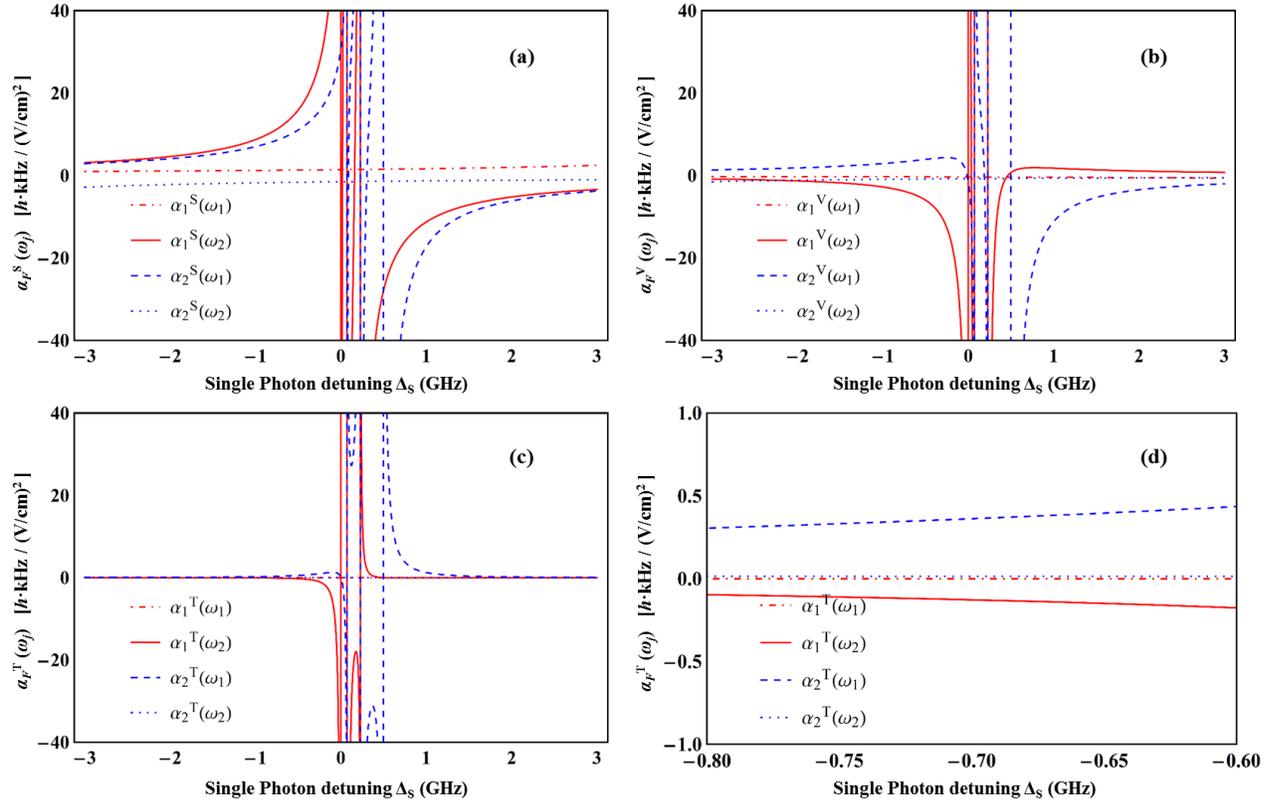

Fig. 2. (a) Scalar, (b) vector, (c) tensor and (d) zoomed in tensor polarizabilities of the $^{87}$Rb atom $F=1$ (red color) and $F=2$ (blue color) ground states. For each sub-graph, the polarizabilities induced by the near-resonant lasers have an apparent difference, and are much larger than the polarizabilities induced by the far-detuned Raman lasers. Specifically, the polarizabilities $\alpha_1^l(\omega_2)$ (red solid lines) and $\alpha_2^l(\omega_1)$ (blue dash lines) with single photon detuning $|\Delta_s| \leq 2\,\text{GHz}$ are much larger than the polarizabilities $\alpha_1^l(\omega_1)$ (red dash dot

lines) and $\alpha_2^l(\omega_2)$ (blue dot lines) with single photon detuning equals $\Delta_s + 6.8$ GHz.

Table 1. List of the dynamic polarizabilities for $^{87}$Rb atom ground states induced by Raman laser field $\mathbf{E}_1(\omega_1)$ and $\mathbf{E}_2(\omega_2)$ with frequencies $\omega_1 = \omega_{12} - 700\,\text{MHz}$, $\omega_2 = \omega_1 + 6.834\,\text{GHz}$.

| Type | Atom state | Dynamic polarizability in $h \cdot$kHz/(V/cm)$^2$ | |
|---|---|---|---|
| | | Raman laser $\mathbf{E}_1(\omega_1)$ | Raman laser $\mathbf{E}_2(\omega_2)$ |
| Scalar $\alpha_F^S(\omega)$ | F=1 | 1.28505 | 13.13140 |
| | F=2 | 9.61215 | -1.68227 |
| Vector $\alpha_F^V(\omega)$ | F=1 | -0.33489 | -4.68128 |
| | F=2 | 3.60581 | -0.87198 |
| Tensor $\alpha_F^T(\omega)$ | F=1 | -0.00099 | -0.16109 |
| | F=2 | 0.41225 | 0.01315 |

### C. Differential ac Stark shift on Raman transition

The differential ac stark shift on Raman transition can be decomposed into scalar, vector, and tensor parts as:

$$\Delta E^{ac} = \Delta E^S + \Delta E^V + \Delta E^T \tag{11}$$

where $\Delta E^S$, $\Delta E^V$ and $\Delta E^T$ are inferred by first summing the light shift on $F=2$ and $F=1$ states induced by the two Raman lasers, respectively, and then calculating the difference as:

$$\Delta E^S = -\left(\frac{\varepsilon_2}{2}\right)^2 \left\{ \left[q\alpha_2^S(\omega_1) + \alpha_2^S(\omega_2)\right] - \left[q\alpha_1^S(\omega_1) + \alpha_1^S(\omega_2)\right] \right\} \tag{12}$$

$$\Delta E^V = -\left(\frac{\varepsilon_2}{2}\right)^2 (\hat{k} \cdot \hat{B}) \Delta m_F \left\{ \frac{1}{4}\left[q\alpha_2^V(\omega_1) + \alpha_2^V(\omega_2)\right] - \frac{1}{2}\left[q\alpha_1^V(\omega_1) + \alpha_1^V(\omega_2)\right] \right\} \tag{13}$$

$$\Delta E^T = -\left(\frac{\varepsilon_2}{2}\right)^2 \left(3|\hat{\zeta} \cdot \hat{B}|^2 - 1\right) \left\{ \frac{3m_F^2 - 6}{12}\left[q\alpha_2^T(\omega_1) + \alpha_2^T(\omega_2)\right] - \frac{3m_F^2 - 2}{2}\left[q\alpha_1^T(\omega_1) + \alpha_1^T(\omega_2)\right] \right\} \tag{14}$$

where $\varepsilon_2$ is the amplitude of the laser field $\mathbf{E}_2(\omega_2)$, $q = I_1/I_2$ is the intensity ratio between Raman laser $\mathbf{E}_1(\omega_1)$ and $\mathbf{E}_2(\omega_2)$, and $\alpha_F^l(\omega_j)$ represents the scalar ($l=S$), vector ($l=V$), and tensor ($l=T$) polarizabilities of $^{87}$Rb atom on $F=1$ state and $F=2$ state induced by Raman lasers $\mathbf{E}_1(\omega_1)$ and $\mathbf{E}_2(\omega_2)$ calculated from Eqs. (8) - (10), respectively.

### III. Experiments and results
#### A. Experimental setup and procedure

The experimental setup used for measuring the light shifts is based on the gravimetric atom interferometer (GAIN) [26,34] and schematically shown in Fig. 3. First, $^{87}$Rb atoms are cooled and trapped in the magnetically shielded magneto-optical trap (MOT) chamber within 0.6 s, and then launched vertically by moving molasses technique, achieving a temperature of ~2 μK and an initial velocity of ~4.4 m/s. The repumping laser is switched off 1 ms later than the cooling laser to ensure all atoms are in the $F=2$ ground state. Next, 0.15 s after launch, a pair of circularly polarized ($\sigma^+\sigma^+$ or $\sigma^-\sigma^-$) Raman π pulse with an e$^{-2}$ diameter of 29.5 mm are irradiated from the top vacuum window when the atoms are moving in the magnetically shielded interferometer chamber. The background magnetic field inside the chamber is attenuated by 30 dB and the residual magnetic field fluctuation was expected to be less than 2 nT. A solenoid, precisely wound inside the magnetic shield, is driven by a precision laser diode current driver (the RMS current noise is ~1 μA) with a current of 6.5 mA to create a highly homogeneous and stable quantization magnetic field of 2936.4 nT in vertical direction [35]. Last, when the atoms fall down through the detection chamber, a normalized fluorescence detection process is applied to obtain the atomic transition probability from $F=2$ state to $F=1$ state, denoted as $P_1$. By repeating the above launch-detection process and sweeping the relative frequency of the two Raman lasers through a DDS frequency controller, three transition peaks will be observed (as shown in Fig.4) and the peak frequency of the $|F=2, m_F\rangle \leftrightarrow |F=1, m_F\rangle$ ($m_F = 0, \pm 1$) transition, denoted as $\omega_{p_{m_F}}$, can be acquired by fitting the Raman spectrum with [36]:

$$P_1 = \frac{\Omega_{eff}^2}{\Omega_{eff}^2 + (\omega - \omega_{p_{m_F}})^2} \sin^2\left[\tau \sqrt{\Omega_{eff}^2 + (\omega - \omega_{p_{m_F}})^2}\right] \quad (15)$$

where $\Omega_{eff} = \Omega_1^* \Omega_2 / 4\Delta_s$ is the effective Rabi frequency, $\Omega_i$ ($i=1, 2$) is the Rabi frequency used to characterize the level coupling by the $i$th Raman beam, $\tau$ is the Raman pulse duration, $\omega_{p_{m_F}} = \Delta E_{Zeeman}/\hbar + \Delta E^{ac}/\hbar + \delta_D$ includes the Zeeman shift $\Delta E_{Zeeman}/\hbar$, the differential ac Stark shift $\Delta E^{ac}/\hbar$, and the Doppler frequency shift $\delta_D$.

Considering Eq. (15), the Zeeman shift $\Delta E_{Zeeman}/\hbar = \pm 41.109$ kHz is a constant, the Doppler shift $\delta_D$ is very small (223.7 mHz during an 1 ms Raman pulse) compared to the ac stark shift (several kHz), we can observe the ac Stark shift from the peak frequencies of the Raman spectra for different laser intensities while maintaining the Raman π pulse condition. The laser intensity is controlled by the working voltage applied to the RF attenuator for Raman laser AOM with a precisely known voltage-to-power proportional relationship and determined by the effective Rabi frequency (Fourier-width) of the acquired Raman spectrum (see Fig. 4). Afterwards, the influence of Raman laser polarization is tracked by repeating this set of experiments with a flipped polarization direction.

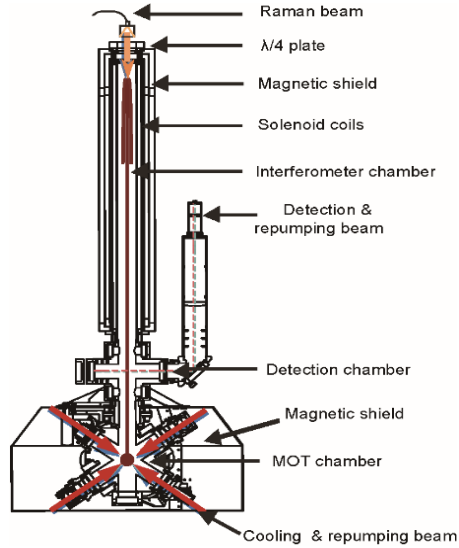

Fig. 3. Schematic diagram of the experimental setup.

## B. Total ac stark shifts

The achieved Raman spectra in $\sigma^+\sigma^+$ and $\sigma^-\sigma^-$ polarization configurations are shown in Figs. 4(a) and 4(b) respectively, where the horizontal axes represent the relative frequencies of the two Raman lasers minus the ground state hyperfine splitting $\omega_{hfs}$. It is clear that with the increasing laser intensity for the Raman π pulse, the left and right peaks of the Raman spectra show an increasing shift in opposite directions due to the vector light shift as shown in Eq. (13) and detailed in Ref. [35]. The acquired Raman spectra also show a broadening effect due to the increasing Fourier width of Raman pulse. The peak frequencies in $\sigma^+\sigma^+$ and $\sigma^-\sigma^-$ polarization configurations obtained by fitting the Raman spectra with a combined Raman transition function of Eq. (15) (curves in Fig. 4), denoted as $\omega_{p_{m_F}}$ and $\omega'_{p_{m_F}}$ ($m_F = 0, \pm 1$), are plotted as points in Fig. 5.

Considering the propagation directions of the Raman lasers are parallel to the quantization magnetic field in our experiment, here $\hat{k} \cdot \hat{B}=1$ and $|\hat{\zeta} \cdot \hat{B}|^2 = 0$ respectively. Substituting the laser frequencies $\omega_1 = \omega_{12} - 700\,\text{MHz}$, $\omega_2 = \omega_1 + 6.834\,\text{GHz}$ into Eqs. (8)-(14), the theoretically inferred differential light shift of the $|F=2, m_F\rangle$ to $|F=1, m_F\rangle$ transition is:

$$\Delta E^{ac} = \varepsilon_2^2 h\left[3.66 - 2.13q - m_F A(0.53 + 0.27q) + m_F^2(0.06 + 0.03q)\right] \quad (16)$$

where $\varepsilon_2$ is the amplitude of the laser field $E_2(\omega_2)$, $h$ is the Planck's constant. $\Delta E^{ac}$ is in the unit of kJ when $\varepsilon_2$ is in the unit of V/cm. The intensity ratio $q$ is set to 1.72 in order to cancel the differential light shift of the $|F=2, m_F=0\rangle$ to $|F=1, m_F=0\rangle$ transition [26]. The Zeeman shifts for $m_F=\pm 1$ branches are $\Delta E_{\text{Zeeman}}/\hbar = \pm 41.109\,\text{kHz}$ respectively. Substituting the above parameters into Eqs. (1) and (16), we can infer the peak frequencies for the $\sigma^-\sigma^-$ and $\sigma^+\sigma^+$ polarization configurations shown as solid and dashed lines in Fig. 5, respectively.

Both the theoretical (lines) and experimental results (points) in Fig. 5 show that the differential light shift of the $|F=2, m_F=0\rangle$ to $|F=1, m_F=0\rangle$ transition is zero, the light shifts of the $|F=2, m_F\rangle$ to $|F=1, m_F\rangle$ ($m_F=\pm 1$) transitions are polarization dependent and proportional to the laser intensities which induce them in the first place.

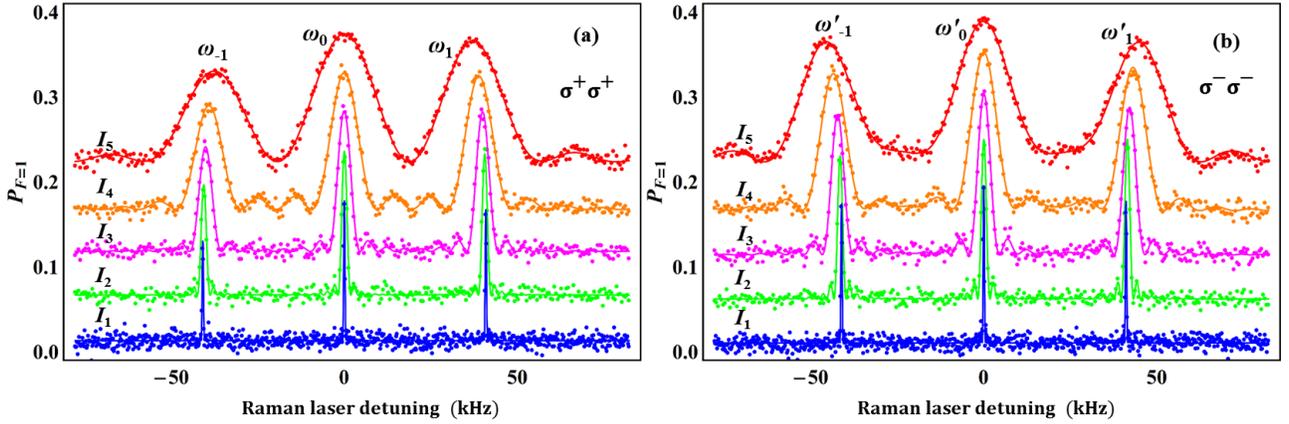

Fig. 4. Raman spectra obtained with different Raman π pulse intensities for the (a) $\sigma^+\sigma^+$ and (b) $\sigma^-\sigma^-$ polarization configurations. The laser intensities $I_1, I_2, I_3, I_4, I_5$ in the plots are 0.241, 0.480, 1.196, 2.493, 4.784 mW/cm², respectively.

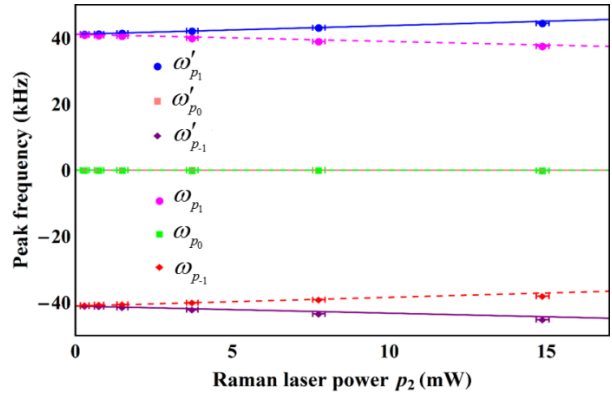

Fig. 5. The dependence of the transition peak frequencies $\omega_{p_{m_F}}$ and $\omega'_{p_{m_F}}$ ($m_F = 0, \pm 1$) on Raman laser intensity in $\sigma^+\sigma^+$ (dashed lines) and $\sigma^-\sigma^-$ (solid lines) polarization configurations. The experimentally measured results (points) are in good agreement with the theoretically calculated results (lines). The horizontal uncertainties come from the power fluctuations of the laser in 1 minute measured with a power meter and the vertical uncertainties come from the fitting uncertainties.

## C. Differential vector and tensor light shifts

From Eqs. (2) and (4), we can see that for a particular setting of Raman laser power and polarization, the differential vector light shift $\Delta\omega_{m_F}^V$ can be extracted by:

$$\Delta\omega_{m_F}^V = \frac{1}{2}\left(\omega_{p_{m_F}} - \omega'_{p_{m_F}}\right) \quad (17)$$

where $\omega_{p_{m_F}}$ and $\omega'_{p_{m_F}}$ ($m_F = 0, \pm 1$) represent the peak frequencies of the $|F=2, m_F\rangle$ to $|F=1, m_F\rangle$ Raman transitions obtained from Raman spectra in $\sigma^+\sigma^+$ (Fig. 4(a)) and $\sigma^-\sigma^-$ (Fig. 4(b)) polarization configurations, respectively. The Zeeman shift, scalar and tensor light shifts for both peak frequencies are equal and canceled in the subtraction. As shown in Fig. 6(a), the experimentally measured differential vector light shifts $\Delta\omega_{m_F}^V$ ($m_F = -1$, 0, 1, shown as magenta rhombuses, red squares and blue circles respectively) for different Raman laser powers match with the theoretically calculated results of the $|F=2, m_F\rangle$ to $|F=1, m_F\rangle$ transitions ($m_F = -1$, 0, 1, shown as magenta dot-dashed line, red dashed line and blue solid line respectively).

The differential tensor light shift, which are equal for the four magnetically sensitive transition peak frequencies $\omega_{p_{-1}}$, $\omega_{p_1}$, $\omega'_{p_{-1}}$ and $\omega'_{p_1}$, only can be extracted by subtracting the Zeeman shift $\Delta\omega_{\text{Zeeman}}$, the differential scalar light shift $\Delta\omega^S$, and the differential vector light shift $\Delta\omega_{m_F}^V$ successively. Therefore, we can only infer the difference between the differential tensor light shifts by the formula:

$$\Delta\omega_1^T - \Delta\omega_0^T = \Delta\omega_{-1}^T - \Delta\omega_0^T = \frac{1}{2}\left[\left(\omega_{p_1} - \Delta\omega_{\text{Zeeman}} - \omega_{p_0}\right) + \left(\omega_{p_{-1}} + \Delta\omega_{\text{Zeeman}} - \omega_{p_0}\right)\right] \quad (18)$$

where $\Delta\omega_{-1}^T$, $\Delta\omega_0^T$ and $\Delta\omega_1^T$ represent the differential tensor light shifts for the $|F=2, m_F\rangle$ to $|F=1, m_F\rangle$ transitions with $m_F = -1$, 0, 1, respectively. As shown in Fig. 6(b), the experimentally measured difference between the differential tensor light shifts of $\Delta\omega_{1(-1)}^T$ and $\Delta\omega_0^T$ in the $\sigma^+\sigma^+$ (blue circles) and $\sigma^-\sigma^-$ (red squares) polarization configurations match with the theoretically calculated results (magenta solid line).

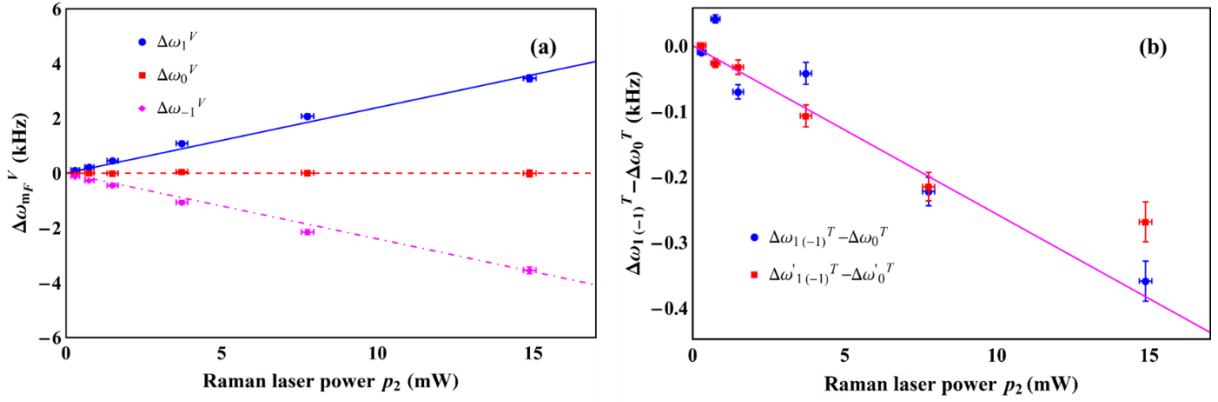

Fig. 6 The differential vector and tensor light shifts for different Raman laser powers. (a) differential vector light shifts $\Delta\omega_{m_F}^T$, where the measured results for $m_F = -1, 0, 1$ are denoted as magenta rhombuses, red squares and blue circles respectively, the theoretically inferred results are shown as magenta dot-dashed line, red dashed line and blue solid line, respectively. (b) difference between the differential tensor light shifts, where the measured results $\Delta\omega_{1(-1)}^T - \Delta\omega_0^T$ for $\sigma^+\sigma^+$ and $\sigma^-\sigma^-$ polarization configurations are denoted as blue circles and red squares respectively, the theoretically inferred results are shown as magenta solid line. The horizontal uncertainties come from the power fluctuations of the laser in 1 minute measured with a power meter and the vertical uncertainties come from the fitting uncertainties, see Ref. [35] for detail.

## IV. Conclusions and outlook

In conclusion, we have presented the derivation of the frequency dependent dynamical polarizabilities and differential light shifts of the $^{87}$Rb atom 5s ground state sub-levels interacting with a pair of near-resonant Raman light fields. Using a simple method of stimulated Raman spectroscopy, the measured differential vector and tensor light shifts for Raman detuning of -700 MHz match with the calculations. For $^{87}$Rb atom $F=1$ and $F=2$ ground states, we experimentally verified vector dynamical polarizabilities of -4.68128 $h\cdot$kHz/(V/cm)$^2$ and 3.60581 $h\cdot$kHz/(V/cm)$^2$, and tensor dynamical polarizabilities of -0.16109 $h\cdot$kHz/(V/cm)$^2$ and 0.41225 $h\cdot$kHz/(V/cm)$^2$, respectively.

We further clarify the common concept that the tensor polarizability for the ground state of alkali atoms is absent with conditions that the light field is far-detuned from resonance with the atom and the total electronic angular momentum J is a good quantum number. Otherwise, the light shifts are non-trivial and will shift the resonant frequencies of the Raman transitions. The characterization of the frequency dependent dynamic polarizabilities will help to determine the intensity ratio or frequency of the laser in order to extinct or stabilize the light shifts in arbitrary laser polarization, detuning and atomic magnetic states. Vice versa, this work is helpful for evaluation the laser induced scalar, vector and tensor light shifts and subtracting them in the data analysis process for a given intensity ratio and frequency of the laser, such as subtracting the vector light shift induced fictitious magnetic field in atomic magnetometers.

**Acknowledgments**

This material is based on work funded by the European Commission (FINAQS, Contract. No. 012986-2 NEST), by ESA (SAI, Contract. No. 20578/07/NL/VJ) and by ESF/DFG (EuroQUASAR-IQS, DFG grant PE 904/2-1 and PE 904/4-1). Q. Q. Hu would like to thank the support of the National Natural Science Foundation of China under Grant No. 51275523, Specialized Research Fund for the Doctoral Program of Higher Education of China under Grant No. 20134307110009, and the Graduate Innovative Research Fund of Hunan Province under Grant No. CX2014A002.